\documentclass[11pt,a4paper]{article}
\usepackage{jcappub}

\usepackage{amsmath, amsfonts,amsthm, amssymb, mathrsfs, bm, verbatim, pslatex, bigints, physics, tikz}
\usetikzlibrary{decorations.markings}

\def\be{\begin{equation}}
\def\ee{\end{equation}}
\def\bea{\begin{eqnarray}}
\def\eea{\end{eqnarray}}
\def\bml{\begin{subequations}}
\def\blea{\bml\begin{eqnarray}}
\def\elea{\end{eqnarray}\end{subequations}}

\usepackage{amsmath}

\begin{document}

\title{Generating Functionals for Quantum Field Theories with Random Potentials}

\author{Mudit Jain and}

\emailAdd{jainx286@d.umn.edu}

\author{Vitaly Vanchurin}

\emailAdd{vvanchur@umn.edu}

\date{\today}

\affiliation{Department of Physics, University of Minnesota, Duluth, Minnesota, 55812}

\abstract{

We consider generating functionals for computing correlators in quantum field theories with random potentials.  Examples of such theories  include cosmological systems in context of the string theory landscape (e.g. cosmic inflation) or condensed matter systems with quenched disorder (e.g. spin glass). We use the so-called replica trick to define two different generating functionals for calculating correlators of the quantum fields averaged over a given distribution of random potentials. The first generating functional is appropriate for calculating averaged (in-out) amplitudes and involves a single replica of fields, but the replica limit is taken to an (unphysical) negative one number of fields outside of the path integral. When the number of replicas is doubled the generating functional can also be used for calculating averaged probabilities (squared amplitudes) using the in-in construction. The second generating functional involves an infinite number of replicas, but can be used for calculating both in-out and in-in correlators and the replica limits are taken to only a zero number of fields. We discuss the formalism in details for a single real scalar field, but the generalization to more fields or to different types of fields is straightforward. We work out three examples: one where the mass of scalar field is treated as a random variable and two where the functional form of interactions is random, one described by a Gaussian random field  and the other by a Euclidean action in the field configuration space.
 
}

\maketitle

\section{Introduction}

The standard assumption in quantum field theories is that the quantum action is some fixed functional of quantum fields. Similarly in statistical mechanics it is usually assumed that the Hamiltonian is some fixed functional of random fields and its conjugate momenta. In both cases the randomness in experimental outcomes may only come from the fact that the (quantum and/or random) fields are described by distributions encoded entirely in the (quantum and/or statistical) partition functionals, but there is no randomness in the theory itself (e.g.  Lagrangian or Hamiltonian  densities are fixed). This is, perhaps, a good approach if one is allowed to perform (and compare against each other) a large number of experiments so that the initial randomness in a theory (due to the fact that the theory is not known a priory) becomes negligible.\footnote{Note, that the initial uncertainties would never get completely eliminated, but they can be made arbitrary small with more and more data coming from experiments.} There are, however, at least two cases when the uncertainty in the theory cannot be reduced beyond a certain limit and thus the assumption of a fixed theory should not be used. The first case is when the theory contains {\it too many} parameters (describing for example a very complicated potential) all of which cannot be fixed simultaneously due to experimental limitations. The second case is when there are {\it too few} experiments that one can perform on the system (due to, for example, cosmic variance). The first case is usually studied in the context of disordered systems in condensed matter and the second case is present in cosmology, since there is only one universe to observe. For example, it was known for some time that a particular distribution of random potentials gives rise to the phenomena of Anderson localization \cite{6_Cardy, 7_BDFN, 4_Anderson:1958vr, 5_Kamenev}, but more recently there has also been a quest to understand quantum inflationary perturbations on random potentials  \cite{8_Tegmark:2004qd, 9_Denef:2004cf, 10_Marsh:2011aa}. The situation is complicated further in the context of string theory models of eternal inflation, where one has to deal with both cases simultaneously: a large number of parameters (of the string theory landscape \cite{1_Bousso:2000xa, 2_Kachru:2003aw, 3_Susskind:2003kw}) and only a single observational event of the Universe we live in.\footnote{Strictly speaking the models of eternal inflation also suffer from an additional uncertainty of choosing a probability measure \cite{Vanchurin:1999iv} (known as the measure problem) which will be discussed in a follow-up publication \cite{33_Mudit_Vitaly}.}  To determine the statistical properties in such systems one usually generates numerically a single random potential and then simulates directly the classical background dynamics \cite{11_Bachlechner:2014rqa, 12_Chen:2011ac, 13_Bachlechner:2012at, 14_Battefeld:2008qg, 15_Battefeld:2014qoa, 16_Frazer:2011tg, 17_Frazer:2011br}. In this work we present a fully analytical (and also quantum) treatment of random potentials using the framework of path integrals by defining two generating functionals both of which can in principle be used for calculating averaged correlators. The analysis is done for a single real scalar quantum field theory with a distribution in potentials, but the generalization to many fields and to different types of fields is straightforward. 

By random potential we mean that the potential had been drawn from a statistical ensemble described by some probability distribution $P[V]$ which is known, but the exact realization of the potential $V(\varphi)$ is unknown. The simplest case perhaps is if there are no (non-quadratic) interactions, i.e. $V(\varphi)=\frac{1}{2}m^2\varphi^2$ and all of the randomness is imprinted in the value of the mass $m$, chosen according to some distribution $P(m)$. A distribution in the coupling constant of the non-quadratic interactions may be chosen as well. A more interesting case though might be to consider a functional distribution for the potential which can be described by some (Euclidean) action in field's configuration space, 
\be
P[V] = \frac{1}{N} \exp\left(-\frac{1}{\hbar_V} \int d\varphi \;{\cal L}\left (\frac{dV(\varphi)}{d\varphi}, V(\varphi)\right )\right) \label{eq:V_distribution}
\ee
with
\be
\int DV\,P[V] =1. \label{eq:normalization}
\ee
where $\hbar_V$ is an arbitrary constant in the distribution $P[V]$ that need not be related to Planck's constant $\hbar=1$.  

Note that the potential `field' $V(\varphi)$ lives in the configuration field space and not in physical space-time and the expression of Eqs. \eqref{eq:V_distribution} and \eqref{eq:normalization} is nothing but a Euclidean path integral of `field' $V$ with `time' variable $\varphi$. A discretized version of Eq. \eqref{eq:V_distribution} is understood as 
\be
P[V] = \frac{1}{N} \exp\left(-\frac{1}{\hbar_V} \sum_i \epsilon \;{\cal L}\left (\frac{V(\varphi_{i+1}) - V(\varphi_{i})}{\epsilon}, V(\varphi_i)\right)\right)
\ee
where $\epsilon$ is the discretization scale of `time' variable $\varphi$, $V(\varphi_i)$ are the values of `field' at discrete points $\varphi_i$,  and the measure of path integration in \eqref{eq:normalization} is understood as 
\be
DV = \prod_i dV_i.
\ee This in turn will make the integral in the path integral over field (i.e. $\int D\varphi f[\varphi] = \int \prod_j d \varphi_j f(\varphi_0, \varphi_1, ...)$) to be rather a sum
\be
\int D\varphi f[\varphi]  = \sum_{i_0} \sum_{i_1} \sum_{i_2} ....  f(\varphi_{i_0}, \varphi_{i_1}, \varphi_{i_2}, ...)
\ee
where $f[\varphi]$ is an arbitrary functional of function of space-time $\varphi(x)$. This shows how the discretization can be carried out in principle, but we will attempt to work in a continuum limit for the remainder of the paper. 

As discussed in Sec.  \ref{Sec:HarmonicOscillator}, having `Lagrangians' for potential described by Eqs. \eqref{eq:V_distribution} and \eqref{eq:normalization} leads to a technical issue and analytical calculations seems formidable, but one can still tackle the problem numerically. More generally, the statistical ensemble of random potentials need not be described by a local action as in \eqref{eq:V_distribution}, for instance it could be described by some Gaussian random field with a prescribed two point correlator $\Delta_{V}$ for the potential, i.e.
\be
P[V] = \dfrac{1}{N}\exp\left(-\iint d\varphi_1\,d\varphi_2\,V(\varphi_1)\Delta^{-1}_V(\varphi_1, \varphi_2)\,V(\varphi_2)\right),
\label{gauss_rand_pot}
\ee
where the integration can be understood by discretizing the field space as was described in the previous paragraph if needed. In either case, the methods of generating functionals developed in this paper apply. 

The properties of quantum (or random) fields such as amplitudes or correlators are conveniently encoded in generating functionals for correlators also known as quantum (or statistical) partition functional. For example, the partition functional for a real quantum scalar field is given by \cite{18_Zee:2003mt, 19_Ryder:1985wq}
\be
Z[V,J] \equiv \int \,D\,\varphi\,\exp\left ( i\,\int d^4 x\left(\frac{1}{2} \partial_{\mu}\varphi \partial^{\mu}\varphi  - V(\varphi) + J\varphi\right)\right) \label{eq:Z}
\ee where we have set $\hbar = 1$. Then the in-out \footnote{For most of this paper, we refer to in and out vacuum states (i.e. $|vac, in \rangle$ and $|vac, out\rangle$) as just in and out states respectively without mentioning the term vacuum, with the hope that it is clear from the context.} correlators can be obtained by functional differentiation with respect to sources at different points, with proper normalization, i.e. 
\be
\langle T\{\varphi(x_k)..\varphi(x_1)\}\rangle = \left. \frac{(-i)^k \delta^k Z[V,J]}{Z[V,0]  \delta{J(x_k)..\delta{J(x_1)}}}\right|_{J = 0} \label{eq:in-out}
\ee
where $T$ stands for time ordering. From now on, we suppress writing out the $T$ operation, but it is always assumed to be present inside correlators. Next, the generating functional for \emph{in-in} correlators \footnote{We refer to in-in as multiplied in-out and out-in correlators/amplitudes, where out-in are nothing but conjugated in-out correlators, i.e. $\langle in, vac|T^{\dagger}\{\varphi...\varphi\}|vac, out\rangle = \langle out, vac|T\{\varphi...\varphi\}|vac, in\rangle^*$. Although these are not exactly the standard in-in objects as usually defined in QFT's \cite{20_Schwinger, 21_Bakshi:1962dv, 22_Bakshi:1963bn} and cosmology \cite{24_Jordan:1986ug, 23_Adshead:2009cb} where a sum over all possible out states is implied, we will still use the term in-in with the understanding that there is only one out state (the vacuum out state).} is obtained by adding one more (conjugated) partition function (i.e. $Z[V,J^+] Z^*[V,J^-]$ and thus having two fields $\varphi^+$ and $\varphi^-$ with respective sources $J^+$ and $J^-$), and the in-in correlators can be obtained by functional differentiation with respect to sources at different points, i.e. 
\be
\langle T\{\varphi(y_k)...\varphi(y_1)\}\rangle^*\langle T\{\varphi(x_k)...\varphi(x_1)\}\rangle  = \left. 
\frac{\delta^{k} Z^*[V,J^-]}{Z^*[V,J^-] \delta{J^-(y_1)}..\delta{J^-(y_k)}}
\frac{ \delta^{k} Z[V,J^+]}{Z[V,J^+] \delta{J^+(x_k)}..\delta{J^+(x_1)}}
\right|_{J^+ = J^- = 0} \label{eq:in-in}
\ee
Note that in \eqref{eq:Z} it is assumed that $V(\varphi)$ is a known potential. The question that we would like to ask in this paper is what should be the generating functional if the exact form of $V(\varphi)$ is unknown or, in other words, if the masses (or any other parameter in the Lagrangian) and/or the functional form of the potential/interactions are random. Note that the same question can be asked in the context of statistical physics systems where this additional randomness may come from the Hamiltonian itself \cite{18_Zee:2003mt, 6_Cardy, 7_BDFN, 25_Anderson_Edwards}. 

There are four types of averages that we will have to distinguish. Two being statistical in origin (known as annealed and quenched averages in statistical physics \cite{6_Cardy, 7_BDFN}) and two being quantum in origin (in-out and in-in averages/correlators in quantum field theories as recapitulated above). The statistical averages will be denoted with a bar ($\overline{O}_a$ for annealed and $\overline{O}_q$ for quenched) and the quantum averages will be denoted with square brackets ($\langle O \rangle$ for in-out and $\langle O \rangle^* \langle O \rangle$ for in-in) as usual, where $O$ is some operator (observable). The distinction between annealed and quenched averaging or between in-out or in-in averages should be clear from context. 

Annealed average must be taken whenever the target field $\varphi$ can backreact on the random field $V$ and both evolve together (for example as in regular interacting field theories with $V$ replaced by another field 'living' in the same physical space-time).\footnote{There are many other places where annealed averages are studied, e.g. spin glasses in condensed matter physics\cite{6_Cardy, 7_BDFN, 25_Anderson_Edwards, 26_Domi_Giard}; quantum brownian motion\cite{18_Zee:2003mt, 27_J.Zinn-Justin, 28_Zanella&Calzetta}, also see \cite{29_MSR}; and non-equilibrium QFT\cite{18_Zee:2003mt, 30_Feynman&Vernon, 31_Cal_Hu, 32_Kamenev}.} Then the averaged in-out correlators, for instance, are given by
\be
\overline{\langle \varphi(x_k)...\varphi(x_1)\rangle}_a = \dfrac{\int DV P[V]\int D\varphi e^{iS}\varphi(x_1)...\varphi(x_k)}{\int DV P[V]\int D\varphi e^{iS}}\label{eq:annealed_io}
\ee
where $k$ denotes what point correlator it is. Note that since both the random potential $V$ and the field $\varphi$ can be dynamic, we simply inserted one more path integral for the random field $V$. Similarly, the corresponding averaged in-in correlators  \eqref{eq:in-in} are 
\be
\overline{\langle \varphi(y_k)...\varphi(y_1)\rangle^*\langle \varphi(x_k)...\varphi(x_1)\rangle}_a = 
\dfrac{\int DV P[V]\int D\varphi e^{-iS}\varphi(y_k)...\varphi(y_1)\int D\varphi e^{iS}\varphi(x_1)...\varphi(x_k) }{\int DV P[V]\int D\varphi e^{-iS}\int D\varphi e^{iS}}\label{eq:annealed_ii}
\ee
On the other hand in the case of quenched systems, the random field is quenched (or frozen) and does not evolve in time. This means that the target field whose evolution depends on the random field does not backreact on the latter, which is the case we will be addressing in this paper. Since our random field manifests itself as a potential $V(\varphi)$ in the Lagrangian for the target field $\varphi$, we want to study field theory on random, but with a given distribution $P[V]$, potentials in order to obtain the averaged correlators. Hence, the quenched averaged in-out correlators are given by
\be
\overline{\langle \varphi(x_k)...\varphi(x_1)\rangle}_q = \int DV P[V]\left(\dfrac{\int D\varphi e^{iS}\varphi(x_1)...\varphi(x_k)}{\int D\varphi e^{iS}}\right)\label{eq:quenched_io}
\ee
and correspondingly the quenched averaged in-in correlators are
\be
\overline{\langle \varphi(y_k)...\varphi(y_1)\rangle^*\langle \varphi(x_k)...\varphi(x_1)\rangle}_q=  \int DV P[V]\left(\dfrac{\int D\varphi e^{-iS}\varphi(y_k)...\varphi(y_1)}{\int D\varphi e^{-iS}}\dfrac{\int D\varphi e^{iS}\varphi(x_1)...\varphi(x_k)}{\int D\varphi e^{iS}}\right)\label{eq:quenched_ii}.
\ee
It is understood that the action $S$ for the field $\varphi$ depends on the potential and therefore we need not bother writing it as a functional of $V$ over and over again. Since in this paper we are interested in quenched averages, we do not use annealed averages from here on, moreover $\overline{O}$ without a subscript always indicates a quenched average.

This paper is organized as follows. In Sec. \ref{Sec:SingleReplica} we summarize the replica trick and define a generating functional for quenched averages with a single (or double) replica of fields for calculating in-out (or in-in) correlators. In Sec. \ref{Sec:RandomPotentials} we use the generating functional for calculating averaged correlators for a distribution of random self-interactions described by either a partition function of a Euclidian harmonic oscillator  (see Sec. \ref{Sec:HarmonicOscillator}) or by two-point correlators of a Gaussian random field (see Sec. \ref{Sec:GaussianField}). In Sec. \ref{Sec:MultipleReplicas}  we construct a generating functional for quenched averages with an infinite number of replica of fields which we use in Sec. \ref{Sec:RandomMasses} to calculate averaged correlators for a particular distribution of random masses. The main results of the paper are summarized and discussed in Sec. \ref{Sec:Discussion}.

\section{Single (Double) Replica(s) of Fields}\label{Sec:SingleReplica}

The main challenge in calculating quenched averages is the dependence on random potentials in the denominators in expressions \eqref{eq:quenched_io} and  \eqref{eq:quenched_ii}. The problem is usually handled by either the so-called replica trick or by introducing Grassmanian fields, see e.g. the study of disordered systems in condensed matter \cite{18_Zee:2003mt, 6_Cardy, 7_BDFN, 5_Kamenev, 25_Anderson_Edwards}. However one major difference from those condensed matter studies is that there the potential `lives' in the same physical space where the field `lives', whereas in our case, which is relevant for cosmological systems and in general for relativistic systems, the potential `lives' in the field space of $\varphi$. Also, we want to obtain a generating functional to generate all averaged correlators, all of which are potentially observable quantities. In what follows we present two generating functionals for generating averaged correlators, with one (see Sec. \ref{Sec:MultipleReplicas}) being superior and more trustworthy compared to another in the sense that only interpolation of the functional between positive integers is needed and not extrapolation to negative integers (discussed later), but at the expense of increased complexity. The formalism is illustrated for a real scalar field with random potentials (see Sec. \ref{Sec:RandomPotentials}), and random masses (see Sec. \ref{Sec:RandomMasses}), and we note that the generalization to different types and collections of fields is straightforward. Possible application of the proposed recipe to inflationary cosmology will be discussed in detail in a separate publication \cite{33_Mudit_Vitaly}.

To illustrate the replica trick we can make use of a trivial identity
\be
\frac{1}{Z[V,J]} = \lim_{n \rightarrow -1} Z[V,J]^n,
\ee
where $Z[V,J]$ is the partition functional with a fixed potential. Then generating functionals for quantum systems with random potentials with single and double replica can be defined respectively as
\be
\mathcal{Z}^{\text{io}}[J, \tilde{J}] = \lim_{n\rightarrow -1}\int\,DV\,P[V]\,Z[V,J]\,Z^{n}[V,\tilde{J}]
\label{eq:G1}
\ee
and
\be
\mathcal{Z}^{\text{ii}}[J^+, \tilde{J}^+, J^-, \tilde{J}^-] = \lim_{n^+,n^- \rightarrow -1}\int\,DV\,P[V]Z[V,J^+]Z^{n^+}[V,\tilde{J}^+] Z^*[V,J^-]Z^{*\,n^-}[V,\tilde{J}^-].
\label{eq:G2}
\ee
and the quenched averaged  in-out and in-in correlators can be obtained by differentiating \eqref{eq:G1} and \eqref{eq:G2} with respect to sources, i.e.
\be
\overline{\langle \varphi(x_1)...\varphi(x_k)\rangle} =\left. \frac{ (-i)^k \delta^{k} \mathcal{Z}^{\text{io}}}{\delta J(x_k)..\delta J(x_1)}\right|_{J=\tilde{J} = 0} \label{eq:G1-in-out}
\ee
and
\be
\overline{\langle \varphi(y_1)...\varphi(y_k)\rangle^*\langle \varphi(x_1)...\varphi(x_k)\rangle} = \qquad\left. \dfrac{\delta^{2k} \mathcal{Z}^{\text{ii}}}{\delta J^-(y_k)..\delta J^-(y_1)\delta J^+(x_k)..\delta J^+(x_1)}\right|_{J^{\pm}=\tilde{J}^{\pm} = 0}.\label{eq:G2-in-in}
\ee\footnote{Note that the tilded sources are introduced precisely to handle dependence of $V$ in the denominator, and they are not used for functional differentiation to obtain correlators.}
However, for the last two equations \eqref{eq:G1-in-out} and \eqref{eq:G2-in-in} to be correct we must make two crucial assumptions. First of all we must assume that the integrands of the path integrals in \eqref{eq:G1} and \eqref{eq:G2} as functions of  $n$  and  $n^+, n^-$ respectively can be not only interpolated between integer values, but also extrapolated to negative values. Secondly we must also assume that one can interchange the order of taking limits and path integrations in \eqref{eq:G1} and \eqref{eq:G2}. In Sec. \ref{Sec:MultipleReplicas} we will construct an improved generating functional $\mathcal{W}$ with multiple replicas where the extrapolation assumption can be relaxed by avoiding taking limits to negative numbers of fields.

Moving on, if there is a functional distribution for potentials, e.g. a Lagrangian for self-interactions \eqref{eq:V_distribution}, a closed form integration of the field's partition function is not possible in general; hence, we resort to a perturbative calculation in this case. For this treatment to work, we need to assume weak random interactions (i.e. interactions with a small coupling constant $g$) so that a perturbative treatment makes sense. Further there is no limiting procedure required in \eqref{eq:G1} and \eqref{eq:G1} as $n$'s are equal to negative one with binomial expansion at work. For example, for a real scalar field with action
\be
S[\varphi,V,J] = \int d^4 x\left(\dfrac{1}{2}\left(\partial_{\mu}\varphi\right)^2 - \frac{1}{2}\,m^2\varphi^2 + gV(\varphi) + J\varphi\right),
\label{eq: action_phi}
\ee
the partition functional can be expanded perturbatively in $g$ as
\be
Z[V,J] = \left[\sum_l \dfrac{\left(i g\int V\left(\frac{\delta}{ i\delta J(z)}\right)d^4 z\right)^l}{l!}\right]Z_{f}[J]
\ee
around the free partition function
\be
Z_{f}[J] = \exp \left(\frac{i}{2}\iint J(u)\Delta_{F}(u-v)J(v)d^4 u\,d^4 v\right),
\ee
Here, $\Delta_{F} (x - y)$ is the free field propagator (negative of the inverse of the free field operator), also called Feynman Green's function, and it is equal to
\be
\Delta_{F} (x) = \lim_{\epsilon \rightarrow 0} \dfrac{1}{(2\pi)^4}\int d^4k \dfrac{e^{i\,k^{\mu}x_{\mu}}}{k^2 - m^2 + i\epsilon} = \dfrac{-i}{(2\pi)^3}\int\left[\Theta(x^0)\dfrac{e^{-i\left(\omega_k\,x^0 - \vec{k}.\vec{x}\right)}}{2\omega_k} + \Theta(-x^0)\dfrac{e^{i\left(\omega_k\,x^0 - \vec{k}.\vec{x}\right)}}{2\omega_k}\right]d^3k
\label{eq: Feyn_propa}
\ee
where the energies $\omega_k = +\sqrt{\vec{k}^2 + m^2}$ are assumed to be positive and and we are using diag$(+,-,-,-)$ as the Minkowski signature. The conjugated propagator (in the conjugated partition function) would be, apart from a negative sign, reversed in time:
\be
\Delta_{F}^* (x) = \dfrac{i}{(2\pi)^3}\int\left[\Theta(x^0)\dfrac{e^{i\left(\omega_k\,x^0 - \vec{k}.\vec{x}\right)}}{2\omega_k} + \Theta(-x^0)\dfrac{e^{-i\left(\omega_k\,x^0 - \vec{k}.\vec{x}\right)}}{2\omega_k}\right]d^3k
\ee
i.e. the positive and negative frequencies get swapped with respect to positive and negative times. In the following section, we describe how the perturbation theory works for random interactions given by some distribution $P[V]$ for a real scalar field with the action \eqref{eq: action_phi}.

\section{Random Potentials}\label{Sec:RandomPotentials}

As described before, for a perturbative treatment in $g$ (i.e. field $\phi$ with random self-interactions), no limiting procedures are required and all $n$'s in \eqref{eq:G1} and \eqref{eq:G2} can be set to negative one. Then the in-out and in-in partition function can be expanded in $g$ as
\be
\mathcal{Z}^{\text{io}}[J, \tilde{J}] =  \mathcal{Z}^{\text{io}(0)}[J, \tilde{J}] + i g \mathcal{Z}^{\text{io} (1)}[J, \tilde{J}] + (i g)^2 \mathcal{Z}^{\text{io}(2)}[J, \tilde{J}] ... 
\label{eq:Zio_perturb}
\ee
and
\be
\mathcal{Z}^{\text{ii}}[J^+, \tilde{J}^+, J^-, \tilde{J}^-] =  \mathcal{Z}^{\text{ii}(0)}[J^+, \tilde{J}^+, J^-, \tilde{J}^-] + i g \mathcal{Z}^{\text{ii} (1)}[J^+, \tilde{J}^+, J^-, \tilde{J}^-]+  (i g)^2 \mathcal{Z}^{\text{ii}(2)}[J^+, \tilde{J}^+, J^-, \tilde{J}^-] ... \;\;\;\;\;\;\;\; 
\label{eq:Zii_perturb}
\ee
where the zeroth order terms are
\bea
\mathcal{Z}^{\text{io}(0)}[J, \tilde{J}] &= \dfrac{Z_f[J]}{Z_f[\tilde{J}]} \label{eq:Zio_perturb0} \\ 
\mathcal{Z}^{\text{ii}(0)}[J^+, \tilde{J}^+, J^-, \tilde{J}^-] &= \dfrac{Z_f[J^+]Z^*_f[J^-]}{Z_f[\tilde{J}^+]Z^*_f[\tilde{J}^-]} ,
\eea
and the first order terms are
\bea
\mathcal{Z}^{\text{io}(1)}[J, \tilde{J}] = \int DV\,P[V] \int \,d^4 z \left (\dfrac{\int D\varphi\, V[\varphi(z)]e^{i\,S_f[\varphi, J]}}{Z_f[J]} - \dfrac{\int D\tilde{\varphi} \,V[\tilde{\varphi}(z)]e^{i\,S_f[\tilde{\varphi}, \tilde{J}]}}{Z_f[\tilde{J}]}\right) \dfrac{Z_f[J]}{Z_f[\tilde{J}]} \;\;\;\;\;\;\;\;\;\;\\
\mathcal{Z}^{\text{ii}(1)}[J^+, \tilde{J}^+, J^-, \tilde{J}^-] = \int DV\,P[V] \int d^4 z \left ( \dfrac{\int D\varphi^+ \,V[\varphi^+(z)]e^{i\,S_f[\varphi^+, J^+]}}{Z_f[J^+]} -  \dfrac{\int D\tilde{\varphi}^+\,V[\tilde{\varphi}^+(z)]e^{i\,S_f[\tilde{\varphi}^+, \tilde{J}^+]}}{Z_f[\tilde{J}^+]}\right . \notag\\  +   \left . \dfrac{\int D\tilde{\varphi}^-\,V[\tilde{\varphi}^-(z)] e^{-i\,S_f[\tilde{\varphi}^-, \tilde{J}^-]}}{Z^*_f[\tilde{J}^-]} - \dfrac{\int D\varphi^-\,V[\varphi^-(z)]e^{-i\,S_f[\varphi^-, J^-]}}{Z^*_f[J^-]} \right) \dfrac{Z_f[J^+]Z^*_f[J^-]}{Z_f[\tilde{J}^+]Z^*_f[\tilde{J}^-]} \;\;\;\;\;\;\;\;
\eea
where $S_f$ denotes the free action (i.e. \eqref{eq: action_phi} with $g=0$), and so on. By swapping the space-time integral(s) and field path integral(s) with path integral(s) in the potential,\footnote{We assume that for any given potential $V(\varphi)$ drawn from the distribution $P[V]$, the  space-time integral(s) and field path integral(s) converge uniformly and hence the integral(s) can be swapped.} we get correlators in potential as interactions (in such a way that the $k$-th order term in $g$ has $k$-point potential correlators as interactions). For example, we have for the first order terms
\be
\mathcal{Z}^{\text{io} (1)}[J, \tilde{J}] = \int \,d^4 z \left (\dfrac{\int D\varphi\, \overline{V[\varphi(z)]}e^{i\,S_f[\varphi, J]}}{Z_f[J]} - \dfrac{\int D\tilde{\varphi} \,\overline{V[\tilde{\varphi}(z)]}e^{i\,S_f[\tilde{\varphi}, \tilde{J}]}}{Z_f[\tilde{J}]}\right)\dfrac{Z_f[J]}{Z_f[\tilde{J}]}
\label{eq:RP_Zio}
\ee
and 
\bea
\mathcal{Z}^{\text{ii}(1)}[J^+, \tilde{J}^+, J^-, \tilde{J}^-] = \int d^4 z \left ( \dfrac{\int D\varphi^+ \,\overline{V[\varphi^+(z)]}e^{i\,S_f[\varphi^+, J^+]}}{Z_f[J^+]} -  \dfrac{\int D\tilde{\varphi}^+\,\overline{V[\tilde{\varphi}^+(z)]}e^{i\,S_f[\tilde{\varphi}^+, \tilde{J}^+]}}{Z_f[\tilde{J}^+]}\right . \notag\\  +   \left . \dfrac{\int D\tilde{\varphi}^-\,\overline{V[\tilde{\varphi}^-(z)]}e^{-i\,S_f[\tilde{\varphi}^-, \tilde{J}^-]}}{Z^*_f[\tilde{J}^-]} - \dfrac{\int D\varphi^-\,\overline{V[\varphi^-(z)]}e^{-i\,S_f[\varphi^-, J^-]}}{Z^*_f[J^-]} \right) \dfrac{Z_f[J^+]Z^*_f[J^-]}{Z_f[\tilde{J}^+]Z^*_f[\tilde{J}^-]} \;\;\;\;\;\;\;\;
\label{eq:RP_Zii}
\eea
where any potential correlator is defined as usual
\be
\overline{V(\varphi_1)...V(\varphi_k)} \equiv \int DV\,P[V]\,V(\varphi_1)...V(\varphi_k).
\label{eq:V_corr}
\ee 
Now that all of the interactions are encoded in correlators  of potentials we can carry on the perturbative calculations for a given set of such correlators. In what follows, we describe two simple examples: in the first one, the desired set of correlators arises when the potential is modeled as a random field described by some Lagrangian in the field $\varphi$ configuration space, and in the second one, we consider a Gaussian random potential described by some two point correlator of the potential. In both of our cases the one-point correlator vanishes (which might not be true in general)  and thus we need to expand to second order in $g$ in order to see the effect of random interactions. For the in-out generating functional the second order term is given by\footnote{Prime is introduced for the squared terms in the perturbation series in the denominator: Since each quantity in the perturbation series is an integrated quantity, squaring (or any power in it) must be handled by differentiating between each other.}
\bea
\mathcal{Z}^{\text{io}(2)}[J, \tilde{J}] = \iint d^4z\,d^4z'\left(\dfrac{\int D\varphi\,\overline{V[\varphi(z)]V[\varphi(z')]}e^{i\,S_f[\varphi, J]}}{2\,Z_f[J]} - \dfrac{\int D\tilde{\varphi}\,\overline{V[\tilde{\varphi}(z)]V[\tilde{\varphi}(z')]}e^{i\,S_f[\tilde{\varphi}, \tilde{J}]}}{2\,Z_f[\tilde{J}]} \right.
\label{eq:Zio_perturb2}  \;\;\;\;\;\;\;\;\;\;\;\;\;\;\;\;\\ \notag
- \dfrac{\iint D\varphi\,D\tilde{\varphi}\,\overline{V[\varphi(z)]V[\tilde{\varphi}(z')]}e^{i\,S_f[\varphi, J] + i\,S_f[\tilde{\varphi}, \tilde{J}]}}{Z_f[J]\,Z_f[\tilde{J}]}+ \left.\dfrac{\iint D\tilde{\varphi}\,D\tilde{\varphi}'\,\overline{V[\tilde{\varphi}(z)]V[\tilde{\varphi}'(z')]}e^{i\,S_f[\tilde{\varphi}, \tilde{J}] + i\,S_f[\tilde{\varphi}', \tilde{J}']}}{Z_f[\tilde{J}]\,Z_f[\tilde{J}']}\right) \dfrac{Z_f[J]}{Z_f[\tilde{J}]}
\eea
and a similar (but much messier) second order term can be derived for the in-in generating functional.  We first start with the Lagrangian approach where we identify a technical problem which does not allow the analysis to be carried out analytically.  The problem is due to a discontinues derivative of the correlators at zero and is likely to be present for the simplest Lagrangians, but can be easily handled numerically. Then we perform a sample analytic calculation for Gaussian random potentials with sufficiently smooth two-point correlators.

\subsection{Euclidean Harmonic Oscillator}\label{Sec:HarmonicOscillator}

The partition functional for potentials (not to be confused with the partition functional for fields \eqref{eq:Z}) described by a Lagrangian \eqref{eq:V_distribution} can be obtained by inserting a source $J_V$ coupled to the potential $V$:
\be
Z_V[V, J_V] = \int D V \exp\left(-\frac{1}{\hbar_V} \int d\varphi \left[{\cal L}_V\left (\frac{dV}{d\varphi}, V\right ) + J_V V \right)\right]
\ee
where ${\cal L}_V$ is the Lagrangian for the potential. In the case of a Euclidean harmonic oscillator we have
\be
Z_V[V, J_V] = \int D V \exp\left(-\frac{1}{\hbar_V}\int d\varphi \left[\frac{1}{2} \left(\frac{d V}{d\varphi}\right)^2 + \frac{1}{2}  m_V^2 V^2+ J_V V \right] \right).
\label{eq: euclid_ho}
\ee
Note that a particular limit of this distribution is known to give rise to the phenomenon of Anderson localization when $\hbar_V, m_V \rightarrow \infty$, such that $m^2_V/\hbar_V$ remains finite. Depending on the values of $\hbar_V$ and $m_V$, we have a Gaussian distribution in both the potential and its derivative (centered at zero) at every point in the field space. To carry out the matrix Gaussian integral we make the assumption that the potential dies off faster than $1/\varphi$ on either infinities, so that we can neglect boundary terms. The corresponding Green's function (with a single argument) of the differential operator $\frac{1}{2 \hbar_V}\left ({m_V}^2 - \frac{d^2}{d\varphi^2}\right)$  is given by
\be
\Delta_V(\varphi) = \dfrac{\hbar_V}{\pi}\int dk\,\dfrac{e^{ik\varphi}}{k^2 + {m_V}^2}. 
\label{eq:Greens}
\ee
This integral can be evaluated without any deformations of contours as the poles $(-im_V,im_V)$ are on the imaginary axes. For $\varphi > 0$, we take the upper half plane in the complex $k$ space to close the integration contour with the real axis and for $\varphi < 0$, we take the lower half plane to get
\be
\Delta_V(\varphi) = \dfrac{\hbar_V}{m_V}\left[\Theta(\varphi)e^{-m_V\varphi} + \Theta(-\varphi)e^{m_V\varphi}\right] = \dfrac{\hbar_V}{m_V}e^{-|m_V\varphi|}
\ee
It is straightforward to generate all higher points correlators of the random potential \eqref{eq:V_corr} from the potential's Green's function \eqref{eq:Greens}: all odd point correlators are zero, the two point correlators is simply
\be
\overline{V(\varphi_1)V(\varphi_2)} = \Delta_V(\varphi_1 - \varphi_2),\label{eq:two-points}
\ee
and all higher even-point correlators are obtained from $\eqref{eq:two-points}$ using Wick's theorem \cite{18_Zee:2003mt, 19_Ryder:1985wq}. 

With that, we can now move back to the generating functionals \eqref{eq:Zio_perturb} and \eqref{eq:Zii_perturb}. Retaining only the leading non-zero correction (which is second order in $g$ for the considered example) and replacing the fields as functional derivatives with respect to the respective sources, we get for the in-out generating functional \eqref{eq:Zio_perturb}:
\be
\begin{split}
&\mathcal{Z}^{\text{io}} \approx \dfrac{Z_f}{\tilde{Z}_f}\left[1 + \dfrac{(ig)^2}{2\,Z_f}\iint d^4z\,d^4z'\,\Delta_V\left(\frac{\delta}{i\delta J(z)} - \frac{\delta}{i\delta J(z')}\right)Z_f - \dfrac{(ig)^2}{2\,\tilde{Z}_f}\iint d^4z\,d^4z'\,\Delta_V\left(\frac{\delta}{i\delta \tilde{J}(z)} - \frac{\delta}{i\delta\tilde{J}(z')}\right)\tilde{Z}_f\right.\\
&- \dfrac{(ig)^2}{Z_f\,\tilde{Z}_f}\iint d^4z\,d^4z'\,\Delta_V\left(\frac{\delta}{i\delta J(z)} - \frac{\delta}{i\delta \tilde{J}(z')}\right)Z_f\,\tilde{Z}_f + \left.\dfrac{(ig)^2}{\tilde{Z}_f\,\tilde{Z}'_f}\iint d^4z\,d^4z'\,\Delta_V\left(\frac{\delta}{i\delta \tilde{J}(z)} - \frac{\delta}{i\delta \tilde{J}'(z')}\right)\tilde{Z}_f\,\tilde{Z}'_f\right]
\end{split}
\label{eq:G1_perturb2}
\ee
and a similar expression for the in-in generating functional \eqref{eq:Zii_perturb}. Here we have suppressed the source dependence of the free partition functions, hence the tildes $\tilde{Z_f} \equiv Z_f[\tilde{J}]$ and primes $\tilde{Z_f}' \equiv Z_f[\tilde{J}']$. The correlator has a derivative discontinuity at zero and therefore replacing the fields as functional derivatives w.r.t. sources (as usual in path integral QFT) is problematic at zero. Note that we would be facing the same problem for the in-in generating functional \eqref{eq:Zii_perturb}. This can be dealt with by either modifying the correlator (smooth it at zero) which would modify the Lagrangian (see next section), or by starting with well-behaved correlator(s) which may or may not be described by any Lagrangian. Note that there can still be Lagrangians with sufficiently smooth correlators, but it is not the case for the example of the Euclidean harmonic oscillator considered here.  

\subsection{Gaussian Random Potential}\label{Sec:GaussianField}

Consider a Gaussian random potential defined by the probability distribution
\be
P[V] = \frac{1}{N} \exp\left(-\iint d\varphi_1\,d\varphi_2\,V(\varphi_1)\tilde{\Delta}^{-1}_V(\varphi_1, \varphi_2)\,V(\varphi_2)\right)
\label{eq: modifeuclid_ho}
\ee
whose odd-point correlators vanish and all even-point correlators can be obtained from the two-point correlator $\tilde{\Delta}_V$ using Wick's theorem  \cite{18_Zee:2003mt, 19_Ryder:1985wq}. As an example, we smooth out the correlator \eqref{eq:Greens} using hyperbolic tangent (i.e. replacing $|x|$ with  $x\tanh(\gamma x)$):
\be
\Delta_V(\varphi) \rightarrow \dfrac{\hbar_V}{m_V}\exp(-|m_V|\varphi\tanh(\gamma \varphi)) = \tilde{\Delta}_V(\varphi)
\label{eq:V_correlator}
\ee
with $\gamma$ being a smoothing constant andwe  use this expression in the Gaussian random distribution.\footnote{Note that this correlator might be describable using the Lagrangian formalism of the previous section (with perhaps added correction to the action  \eqref{eq: euclid_ho} that depends on some negative power of $\gamma$ so that in the limit $\gamma \rightarrow \infty$, we recover the Euclidean harmonic oscillator), but the exact form of the Lagrangian is not important for our purposes.} The interactions are still described by the in-out generating functional \eqref{eq:G1_perturb2} with  $\Delta_V$ replaced by $\tilde{\Delta}_V$ and we can calculate Feynman diagrams by expanding the smoothed potential Green's function \eqref{eq:V_correlator} around zero:
\be
\tilde{\Delta}_V(\varphi_1 - \varphi_2) = \dfrac{\hbar_V}{m_V}\left(1 - m_V \,\gamma (\varphi_1 - \varphi_2)^2 + \left(\frac{{m_V}^2\,\gamma^2}{2} + \frac{m_V\,\gamma^3}{3}\right)(\varphi_1 - \varphi_2)^4 + ...\right).
\label{eq:pert_Vcorr}
\ee
Using \eqref{eq:pert_Vcorr}, we approximate \eqref{eq:G1_perturb2} (with smoothed out correlator \eqref{eq:V_correlator}) further by the leading order terms in $\gamma$ which gives rise to quadratic interactions. After all the usual diagram arithmetics \footnote{Since the sourced fields denoted with tildes are never to be differentiated when obtaining averaged field correlators, we can set these sources to zero after calculating respective Feynman diagrams.}, we arrive at
\be
\mathcal{Z}^{\text{io}} = \left[1 + \dfrac{g^2{\hbar_V}\gamma}{2}\iint d^4z\,d^4z' {\left(\begin{tikzpicture}[baseline=(current bounding box.center)]
\draw[gray, thick] (-0.3,0) -- (0.3,0);
\draw[black] (-0.3,0) circle (1pt) node[anchor=east] {$J$};
\filldraw[black] (0.3,0) circle (1pt) node[anchor=west] {$z$};
\end{tikzpicture}\right)
\left(\begin{tikzpicture}[baseline=(current bounding box.center)]
\draw[gray, thick] (-0.3,0) -- (0.3,0);
\draw[black] (-0.3,0) circle (1pt) node[anchor=east] {$J$}; 
\filldraw[black] (0.3,0) circle (1pt) node[anchor=west] {$z'$};
\end{tikzpicture}\right)} + O(g^4)\vphantom{1 + \frac{\sqrt{2\pi}g^2{\hbar_V}^2\,s}{l_V}\iint d^4z\,d^4z'}\right]\dfrac{Z_f[J]}{Z_f[0]}
\ee
where 
\be
\left(\begin{tikzpicture}[baseline=(current bounding box.center)]
\draw[gray, thick] (-0.3,0) -- (0.3,0);
\draw[black] (-0.3,0) circle (1pt) node[anchor=east] {$J$}; 
\filldraw[black] (0.3,0) circle (1pt) node[anchor=west] {$w$};
\end{tikzpicture}\right) \equiv \int d^4u\,J(u)\Delta_{F}(u-w).
\label{eq: sourced_int}
\ee
with solid dots representing internal vertices. The averaged two point correlator is then
\be
\overline{\langle\varphi(x_2)\varphi(x_1)\rangle} = \left.\dfrac{\delta^2\mathcal{Z}^{\text{io}}}{i\delta J(x_2)\,i\delta J(x_1)}\right|_{J = 0} = - i\left(\begin{tikzpicture}[baseline=(current bounding box.center)]
\begin{scope}[very thick,decoration={
    markings,
    mark=at position 0.5 with {\arrow{>}}}
    ] 
\draw[postaction = {decorate}][gray, thick] (-0.3,0) -- (0.3,0);
\draw[black] (-0.3,0) circle (0pt) node[anchor=east] {$x_1$};
\draw[black] (0.3,0) circle (0pt) node[anchor=west] {$x_2$};
\end{scope}
\end{tikzpicture}\right) - g^2{\hbar_V}\,\gamma\left(\begin{tikzpicture}[baseline=(current bounding box.center)]
\begin{scope}[very thick,decoration={
    markings,
    mark=at position 0.5 with {\arrow{>}}}
    ] 
\draw[postaction = {decorate}][gray, thick] (-0.4,0) -- (0.1,0);
\filldraw[black] (-0.4,0) circle (0pt) node[anchor=east] {$x_1$};
\filldraw[black] (0.1,0) circle (1pt) node[anchor=south] {$z$};
\draw[postaction = {decorate}][gray, thick] (0.4,0) -- (0.9,0);
\filldraw[black] (0.4,0) circle (1pt) node[anchor=south] {$z'$};
\filldraw[black] (0.9,0) circle (0pt) node[anchor=west] {$x_2$};
\end{scope}
\end{tikzpicture}\right)
\ee
The first diagram is just the usual propagator
\be
\left(\begin{tikzpicture}[baseline=(current bounding box.center)]
\begin{scope}[very thick,decoration={
    markings,
    mark=at position 0.5 with {\arrow{>}}}
    ] 
\draw[postaction = {decorate}][gray, thick] (-0.3,0) -- (0.3,0);
\draw[black] (-0.3,0) circle (0pt) node[anchor=east] {$x$};
\draw[black] (0.3,0) circle (0pt) node[anchor=west] {$y$};
\end{scope}
\end{tikzpicture}\right) \equiv \Delta_{F}(y-x) 
\label{eq:Diagram1}
\ee
and the second diagram a correction representing the following integral
\be
\left(\begin{tikzpicture}[baseline=(current bounding box.center)]
\begin{scope}[very thick,decoration={
    markings,
    mark=at position 0.5 with {\arrow{>}}}
    ] 
\draw[postaction = {decorate}][gray, thick] (-0.4,0) -- (0.1,0);
\filldraw[black] (-0.4,0) circle (0pt) node[anchor=east] {$x$};
\filldraw[black] (0.1,0) circle (1pt) node[anchor=south] {$w$};
\draw[postaction = {decorate}][gray, thick] (0.4,0) -- (0.9,0);
\filldraw[black] (0.4,0) circle (1pt) node[anchor=south] {$w'$};
\filldraw[black] (0.9,0) circle (0pt) node[anchor=west] {$y$};
\end{scope}
\end{tikzpicture}\right)\equiv \iint d^4w\,d^4w'\Delta_{F}(w - x) \Delta_{F}(y - w') 
\label{eq:Diagram2}
\ee
where arrows specify the direction of time flow (i.e. having usual propagators \eqref{eq: Feyn_propa})\footnote{We choose a convention where time flows from left to right.}. Note that \eqref{eq:Diagram2} is nothing but a multiplication of two diagrams: creation of a particle at $x$ and its annihilation somewhere in between, followed by the creation of a particle somewhere in between and its annihilation at $y$. The integral can be computed easily since the two diagrams (integrals) are disconnected, so that
\be
\left(\begin{tikzpicture}[baseline=(current bounding box.center)]
\begin{scope}[very thick,decoration={
    markings,
    mark=at position 0.5 with {\arrow{>}}}
    ] 
\draw[postaction = {decorate}][gray, thick] (-0.4,0) -- (0.1,0);
\filldraw[black] (-0.4,0) circle (0pt) node[anchor=east] {$x$};
\filldraw[black] (0.1,0) circle (1pt) node[anchor=south] {$w$};
\draw[postaction = {decorate}][gray, thick] (0.4,0) -- (0.9,0);
\filldraw[black] (0.4,0) circle (1pt) node[anchor=south] {$w'$};
\filldraw[black] (0.9,0) circle (0pt) node[anchor=west] {$y$};
\end{scope}
\end{tikzpicture}\right) = \dfrac{1}{m^4} 
\label{eq:Diagram2.2}
\
\ee

A similar calculation for the in-in generating functional \eqref{eq:Zii_perturb} gives 
\be
\begin{split}
\mathcal{Z}^{\text{ii}} = \left[1 + \right.&\left.\dfrac{g^2{\hbar_V}\,\gamma}{2}\iint d^4z_1\,d^4z_2\Bigg\{\left(\begin{tikzpicture}[baseline=(current bounding box.center)]
\draw[gray, thick] (-0.3,0) -- (0.3,0);
\draw[black] (-0.3,0) circle (1pt) node[anchor=east] {$J^+$};
\filldraw[black] (0.3,0) circle (1pt) node[anchor=west] {$z_1$};
\end{tikzpicture}\right)
\left(\begin{tikzpicture}[baseline=(current bounding box.center)]
\draw[gray, thick] (-0.3,0) -- (0.3,0);
\draw[black] (-0.3,0) circle (1pt) node[anchor=east] {$J^+$}; 
\filldraw[black] (0.3,0) circle (1pt) node[anchor=west] {$z_2$};
\end{tikzpicture}\right)+ \left(\begin{tikzpicture}[baseline=(current bounding box.center)]
\draw[gray, thick] (-0.3,0) -- (0.3,0);
\draw[black] (-0.3,0) circle (1pt) node[anchor=east] {$J^-$};
\filldraw[black] (0.3,0) circle (1pt) node[anchor=west] {$z_1$};
\end{tikzpicture}\right)
\left(\begin{tikzpicture}[baseline=(current bounding box.center)]
\draw[gray, thick] (-0.3,0) -- (0.3,0);
\draw[black] (-0.3,0) circle (1pt) node[anchor=east] {$J^-$}; 
\filldraw[black] (0.3,0) circle (1pt) node[anchor=west] {$z_2$};
\end{tikzpicture}\right)\right.\\
& + 2\left(\begin{tikzpicture}[baseline=(current bounding box.center)]
\draw[gray, thick] (-0.3,0) -- (0.3,0);
\draw[black] (-0.3,0) circle (1pt) node[anchor=east] {$J^+$};
\filldraw[black] (0.3,0) circle (1pt) node[anchor=west] {$z_1$};
\end{tikzpicture}\right)
\left(\begin{tikzpicture}[baseline=(current bounding box.center)]
\draw[gray, thick] (-0.3,0) -- (0.3,0);
\draw[black] (-0.3,0) circle (1pt) node[anchor=east] {$J^-$}; 
\filldraw[black] (0.3,0) circle (1pt) node[anchor=west] {$z_2$};
\end{tikzpicture}\right)\Bigg\}+ \left.O(g^4)\vphantom{1 + \frac{\sqrt{2\pi}g^2{\hbar_V}^2\,s}{l_V}\iint d^4z\,d^4z'}\right]\dfrac{Z_f[J^+]\,Z_f[J^-]}{(Z_f[0])^2}
\end{split}
\label{eq:Z2_perturb}
\ee
where the bracketed diagrams with $J^+$ sources are as described before (eq. \eqref{eq: sourced_int}), and $J^-$ sources are the same but with conjugated propagators:
\be
\left(\begin{tikzpicture}[baseline=(current bounding box.center)]
\draw[gray, thick] (-0.3,0) -- (0.3,0);
\draw[black] (-0.3,0) circle (1pt) node[anchor=east] {$J^-$}; 
\filldraw[black] (0.3,0) circle (1pt) node[anchor=west] {$w$};
\end{tikzpicture}\right) \equiv \int d^4u\,J^-(u)\Delta_{F}^*(u-w).
\label{eq: sourced_int2}
\ee
Thus, for example, the averaged two point in-in correlator reads:
\be
\begin{split}
\overline{\langle\varphi(y_2)\varphi(y_1)\rangle^*\langle\varphi(x_2)\varphi(x_1)\rangle}& = \left.\dfrac{(i)^2(-i)^2\delta^4\mathcal{Z}^{\text{ii}}}{\delta J^-(y_2)\,\delta J^-(y_1)\,\delta J^+(x_2)\,\delta J^+(x_1)}\right|_{J^+,J^- = 0}\\
=\qquad\left(\begin{tikzpicture}[baseline=(current bounding box.center)]
\begin{scope}[very thick,decoration={
    markings,
    mark=at position 0.5 with {\arrow{>}}}
    ] 
\draw[postaction = {decorate}][gray, thick] (-0.5,1) -- (0.5,1);
\draw[gray, thick] [postaction = {decorate}](0.5,0.5) -- (-0.5,0.5);
\filldraw[black] (-0.5,1) circle (0pt) node[anchor=east] {$x_1$};
\filldraw[black] (0.5,1) circle (0pt) node[anchor=west] {$x_2$};
\filldraw[black] (-0.5,0.5) circle (0pt) node[anchor=east] {$y_1$};
\filldraw[black] (0.5,0.5) circle (0pt) node[anchor=west] {$y_2$};
\end{scope}
\end{tikzpicture}\right) -  i&g^2{\hbar_V}\,\gamma\left(\left(\begin{tikzpicture}[baseline=(current bounding box.center)]
\begin{scope}[very thick,decoration={
    markings,
    mark=at position 0.5 with {\arrow{>}}}
    ] 
\draw[postaction = {decorate}][gray, thick] (-0.5,0.8) -- (-0.15,0.8);
\draw[postaction = {decorate}][gray, thick] (0.15,0.8) -- (0.5,0.8);
\draw[postaction = {decorate}][gray, thick] (0.5,0.1) -- (-0.5,0.1);
\filldraw[black] (-0.5,0.8) circle (0pt) node[anchor=east] {$x_1$};
\filldraw[black] (0.5,0.8) circle (0pt) node[anchor=west] {$x_2$};
\filldraw[black] (-0.5,0.1) circle (0pt) node[anchor=east] {$y_1$};
\filldraw[black] (0.5,0.1) circle (0pt) node[anchor=west] {$y_2$};
\filldraw[black] (-0.15,0.8) circle (1pt) node[anchor=south] {$z_1$};
\filldraw[black] (0.15,0.8) circle (1pt) node[anchor=south] {$z_2$};
\end{scope}
\end{tikzpicture}\right) - 
\left(\begin{tikzpicture}[baseline=(current bounding box.center)]
\begin{scope}[very thick,decoration={
    markings,
    mark=at position 0.5 with {\arrow{>}}}
    ] 
\draw[postaction = {decorate}][gray, thick] (-0.5,0.4) -- (0.5,0.4);
\draw[postaction = {decorate}][gray, thick] (-0.15,-0.3) -- (-0.5,-0.3);
\draw[postaction = {decorate}][gray, thick] (0.5,-0.3) -- (0.15,-0.3);
\filldraw[black] (-0.5,0.4) circle (0pt) node[anchor=east] {$x_1$};
\filldraw[black] (0.5,0.4) circle (0pt) node[anchor=west] {$x_2$};
\filldraw[black] (-0.5,-0.3) circle (0pt) node[anchor=east] {$y_1$};
\filldraw[black] (0.5,-0.3) circle (0pt) node[anchor=west] {$y_2$};
\filldraw[black] (-0.15,-0.3) circle (1pt) node[anchor=north] {$z_1$};
\filldraw[black] (0.15,-0.3) circle (1pt) node[anchor=north] {$z_2$};
\end{scope}
\end{tikzpicture}
\right)\right)
\end{split}
\ee
where the time reversed flow (arrows from right to left) specifies conjugated propagators i.e.
\be
\left(\begin{tikzpicture}[baseline=(current bounding box.center)]
\begin{scope}[very thick,decoration={
    markings,
    mark=at position 0.5 with {\arrow{>}}}
    ] 
\draw[postaction = {decorate}][gray, thick] (0.3,0) -- (-0.3,0);
\draw[black] (-0.3,0) circle (0pt) node[anchor=east] {$x$};
\draw[black] (0.3,0) circle (0pt) node[anchor=west] {$y$};
\end{scope}
\end{tikzpicture}\right) \equiv \Delta_{F}^*(x-y) 
\label{eq:Diagram1star}
\ee
and
\be
 \left(\begin{tikzpicture}[baseline=(current bounding box.center)]
\begin{scope}[very thick,decoration={
    markings,
    mark=at position 0.5 with {\arrow{>}}}
    ] 
\draw[postaction = {decorate}][gray, thick]  (0.1,0) -- (-0.4,0);
\filldraw[black] (-0.4,0) circle (0pt) node[anchor=east] {$x$};
\filldraw[black] (0.1,0) circle (1pt) node[anchor=south] {$w$};
\draw[postaction = {decorate}][gray, thick] (0.9,0) -- (0.4,0);
\filldraw[black] (0.4,0) circle (1pt) node[anchor=south] {$w'$};
\filldraw[black] (0.9,0) circle (0pt) node[anchor=west] {$y$};
\end{scope}
\end{tikzpicture}\right)\equiv \iint d^4w\,d^4w'\Delta_{F}^*(w'-y) \Delta_{F}^*(x-w) = \dfrac{1}{m^4} 
\label{eq:Diagram2star}
\ee
Also, diagrams appearing one on top of another are nothing but a product of them. Notice that the averaged one point in-in correlator is also non zero due to the third term in \eqref{eq:Z2_perturb} even though all odd-point correlators for potentials vanish, since in totality there are two field operators (causal, anti-causal) which is even, i.e.
\be
\overline{\langle\varphi(y)\rangle^*\langle\varphi(x)\rangle} = \left.\dfrac{(i)(-i)\delta^2\mathcal{Z}^{\text{ii}}}{\delta J^-(y)\,\delta J^+(x)}\right|_{J^+,J^- = 0} = g^2{\hbar_V}\,\gamma
\left(
\begin{tikzpicture}[baseline=(current bounding box.center)]
\begin{scope}[very thick,decoration={
    markings,
   mark=at position 0.5 with {\arrow{>}}}
    ] 
\draw[postaction = {decorate}][gray, thick] (-0.3,0) -- (0.3,0);
\filldraw[black] (-0.3,0) circle (0pt) node[anchor=east] {$x$};
\filldraw[black] (0.3,0) circle (1pt) node[anchor=south] {$z_1$};
\draw[postaction = {decorate}][gray, thick] (0.3,-0.3) -- (-0.3,-0.3);
\filldraw[black] (0.3,-0.3) circle (1pt) node[anchor=north] {$z_2$};
\filldraw[black] (-0.3,-0.3) circle (0pt) node[anchor=east] {$y$};
\end{scope}
\end{tikzpicture}
\right) = \dfrac{g^2{\hbar_V}\,\gamma}{m^4} 
\ee
This suggests that both the normal and time reversed fields interact with each other through the random potential. Since we were doing a perturbative expansion no questionable assumptions (discussed above)  had to be made about the generating functionals \eqref{eq:G1-in-out} and \eqref{eq:G2-in-in}. However for a random distribution of masses $P(m)$ (which we are going to discuss in Sec. \ref{Sec:RandomMasses}) a closed form integration over random potentials is possible, and one might question the necessity of \emph{extrapolating} generating functionals to the unphysical negative one number of fields.  In the next section, we will develop a second generating functional which is a lot more symmetric and, moreover, only needs to be \emph{interpolated} between (a lot more physical) zero and one number of fields.

\section{Multiple Replicas of Fields}\label{Sec:MultipleReplicas}

We start by developing the generating functional for averaged in-out correlators and show how to calculate the corresponding averaged correlators from it. Extension to the in-in generating functional is straightforward.  First note that the issue of having $V$ dependence of partition functions in denominators of \eqref{eq:quenched_io} and \eqref{eq:quenched_ii} can be dealt with by using logarithm which, upon differentiation with respect to the sources produces the required denominator automatically. Hence we consider the average of logarithm of the partition functional (generating functional for connected Feynman diagrams: $W = \ln Z$) 
\be
\overline{W}[J] = \int\,DV\,P[V]W[V,J] = \lim_{n\rightarrow 0^+} \dfrac{d}{dn}\left(\int\,DV\,P[V]Z^{n}[V,J]\right).
\label{eq:Avg_feyndiag}
\ee
The exponentiated object $Z^n$ is well defined at all non-negative integers $n$, but the gain here is that we can study the limit of the (\emph{interpolated}) function at $n\rightarrow 0^+$(i.e. having no fields). In this construction we are avoiding the need for extrapolation to negative values of $n$ which is due to the fact that the object is more symmetric as no distinction is made between partition functions in numerators and denominators in \eqref{eq:quenched_io} and \eqref{eq:quenched_ii}. If \eqref{eq:Avg_feyndiag} is functionally differentiated  with respect to sources, the desired averaged one point in-out correlator is indeed generated
\be
\left.\dfrac{\delta\,\overline{W}}{i\delta\,J(x)}\right|_{J = 0} = \int\,DV\,P[V]\left(\dfrac{\int\,D\,\varphi\,e^{i\,S[\varphi,V]}\,\varphi(x)}{\int\,D\,\varphi\,e^{i\,S[\varphi,V]}}\right) = \overline{\langle\,\varphi(x)\,\rangle},
\label{eq:Avg_io_onepoint}
\ee 
but the generating functional cannot be used to generate all higher point correlators. What it generates though are the \emph{averaged cumulants} of in-out correlators, for instance 
\be
\left.\dfrac{\delta^2\,\overline{W}}{i\delta\,J(x_2)\,i\delta\,J(x_1)}\right|_{J = 0} = \overline{\langle\,\varphi(x_2)\varphi(x_1)\,\rangle} - \overline{\langle\,\varphi(x_2)\rangle\langle\,\varphi(x_1)\rangle}
\label{eq:Avg_io_twopointcum}
\ee
and so on. This is due to the fact that $\overline{\langle\,\varphi(x_2)\rangle\langle\,\varphi(x_1)\rangle} \neq \overline{\langle\,\varphi(x_2)\rangle}\,\,\overline{\langle\,\varphi(x_1)\rangle}$, and thus we are not generating cumulants in the usual sense. 

We could have been expecting this property, since we are averaging only over the connected diagrams, whereas in general, an in-out correlator has both connected and disconnected diagrams. Hence, we need a machinery to build all the missing averaged disconnected diagrams. To accomplish this task, we propose a generating functional with an infinite number of replicated sets of identical fields, each set with a different source\footnote{See \cite{34_Kamenev}, section III wherein two replica sets have been used to calculate the averaged product of two one-point correlator for statistical systems.}:
\be
\mathcal{W}^{\text{io}}(n_1, n_2, ...)[J_1, J_2 ...] = \int\,DV\,P[V]\,e^{n_1W[V,J_1]}e^{n_2W[V,J_2]}... = \int\,DV\,P[V]\,Z^{n_1}[V,J_1]\,Z^{n_2}[V,J_2]... \label{eq:G_infinity}
\ee
The averaged products of different $W$'s are obtained by differentiations with respect to the $n$'s, 
\be
\mathcal{W}^{\text{io}}_{n_1n_2...n_k} \equiv  \overline{W[V,J_1]...W[V,J_k]} =  \left.\dfrac{\partial^p\,\mathcal{W}^{\text{io}}}{\partial\,n_1\,\partial\,n_2...\partial\,n_k}\right|_{n_1, n_2, ...\rightarrow 0^+} 
\ee
where $k$ is an arbitrary integer. These are the averages which will be used for building different correlators. For example, to get the averaged one point correlator we only need to differentiate \eqref{eq:G_infinity} with respect to any one $n$ and the respective source only once, i.e.
\be
\overline{\langle\,\varphi(x)\,\rangle} = \left.\dfrac{\delta\,\mathcal{W}^{\text{io}}_{n_1}}{i\delta\,J_1(x)}\right|_{J_1 = 0} = \mathcal{W}^{\text{io}}_{n_1J_1(x)} 
\ee
which is equivalent to \eqref{eq:Avg_io_onepoint}. For the averaged two point correlator, we need two terms (the second one is needed to cancel out the second term in \eqref{eq:Avg_io_twopointcum}):
\bea
\overline{\langle\,\varphi(x_2) \,\varphi(x_1)\rangle} = \overline{\langle\,\varphi(x_2)\,\varphi(x_1)\rangle}-\overline{\langle\,\varphi(x_2)\rangle\langle\,\varphi(x_1)\rangle} + \overline{\langle\,\varphi(x_2)\rangle\langle\,\varphi(x_1)\rangle} = &\label{eq:G_infinity_twopoint}  \\ \notag
 \left.\dfrac{\delta^2\,\mathcal{W}^{\text{io}}_{n_1}}{i\delta\,J_1(x_2)\,i\delta\,J_1(x_1)}\right|_{J_1 = 0} + \left.\dfrac{\delta^2\,\mathcal{W}^{\text{io}}_{n_1n_2} }{i\delta\,J_2(x_2)\,i\delta\,J_1(x_1)}\right|_{J_1,J_2 = 0} = & \mathcal{W}^{\text{io}}_{n_1J_1(x_1)J_1(x_2)} + \mathcal{W}^{\text{io}}_{n_1n_2J_1(x_1)J_2(x_2)} 
\eea
and so on.

There is a pattern that one might notice in calculating all these averaged correlators so that we can identify simple rules for obtaining them. To understand the origin for these rules it might be helpful to recall from QFT that when the generator for connected diagrams $W$ is functionally differentiated with respect to its source, every of the closed brackets $\langle\,\varphi(x_{l})...\varphi(x_1)\rangle$ after differentiation with respect to source at $x_k$ goes to $\langle\,\varphi(x_k)\varphi(x_{l})...\varphi(x_1)\,\rangle - \langle\,\varphi(x_k)\rangle\langle\varphi(x_{l})...\varphi(x_1)\,\rangle$. Also notice that $\mathcal{W}^{\text{io}}$ differentiated with respect to a source without being differentiated with respect to the corresponding $n$ and evaluated at zero vanishes. Hence, to get an averaged $k$ point in-out correlator, we need to form all \emph{partitions} of the set containing $k$ space-time points $\{x_1, ..., x_k\}$. The total number of partitions is known as Bell's number; each of the partitions corresponds to a term and all such terms have to be added to give the averaged in-out correlator. For example, for $k=3$ the total number of partitions is five: $(\{x_1,x_2, x_3\})$, $(\{x_1,x_2\}, \{x_3\})$,$(\{x_1,x_3\}, \{x_2\})$,$(\{x_2,x_3\}, \{x_1\})$ and  $(\{x_1\},\{x_2\}, \{x_3\})$ and the corresponding terms are $\mathcal{}{W}^{\text{io}}_{n_1\,J_{1}\,J_{1}\,J_{1}}$, $\mathcal{W}^{\text{io}}_{n_1\,n_2\,J_{1}\,J_{1}\,J_{2}}$, $\mathcal{W}^{\text{io}}_{n_1\,n_2\,J_{1}\,J_{2}\,J_{1}}$, $\mathcal{W}^{\text{io}}_{n_1\,n_2\,J_{2}\,J_{1}\,J_{1}}$ and $\mathcal{W}^{\text{io}}_{n_1,\,n_2,\,n_3,\,J_{1},\,J_{2},\,J_{3}}$ respectively. It is assumed that the space-time dependence of the sources is ordered such that the first source depends on $x_1$, the second source depends on $x_2$, etc., e.g. 
\be
\left. \mathcal{W}^{\text{io}}_{n_1\,n_2\,J_{1}\,J_{2}\,J_{1}} \equiv \frac{\partial^2}{\partial n_1\partial n_2} \frac{(-i)^3\delta^3}{\delta J_1(x_3) \delta J_2(x_2) \delta J_1(x_1)}\mathcal{W}^{\text{io}}\right|_{n_1, n_2, ...\rightarrow 0^+, J_1=J_2=...=0}.
\ee
By adding all of the terms we get
\be
\overline{\langle\,\varphi(x_3)\,\varphi(x_2)\,\varphi(x_1)\rangle} = \mathcal{W}^{\text{io}}_{n_1\,J_{1}\,J_{1}\,J_{1}}+ \mathcal{W}^{\text{io}}_{n_1\,n_2\,J_{1}\,J_{1}\,J_{2}} + \mathcal{W}^{\text{io}}_{n_1\,n_2\,J_{1}\,J_{2}\,J_{2}}+ \mathcal{W}^{\text{io}}_{n_1\,n_2\,J_{1}\,J_{2}\,J_{1}} + \mathcal{W}^{\text{io}}_{n_1\,n_2\,n_3\,J_{1}\,J_{2}\,J_{3}}
\ee
So far we have only discussed the one-, two- and three-point correlators, but the same construction of averaged correlators from partitions of the set of space-time coordinates $\{x_1, ..., x_k\}$ works for an arbitrary $k$.

Moving on to the calculation of in-in correlators, the generating functional is a straightforward extension of \eqref{eq:G_infinity}:
\be
\begin{split}
\mathcal{W}^{\text{ii}}(n^{\pm}_{1},.,n^{\pm}_{m})[J^{\pm}_{1},.,J^{\pm}_{m}] = &\int\,DV\,P[V]e^{n^{+}_{1}W[V,J^{+}_1]}...e^{n^+_{m}W[V,J^+_{m}]}e^{n^-_1W^*[V,J^-_1]}...e^{n^-_{m}W^*[V,J^-_{m}]}\\
= &\int\,DV\,P[V]\,Z^{n^+_1}[V,J^+_1]...Z^{*n^-_{m}}[V,J_{m}] \label{eq:G2_infinity}.
\end{split}
\ee
 The two pieces (in-out and out-in) of the averaged in-in correlators  in \eqref{eq:quenched_ii} are obtained from the in-out and out-in integrands of \eqref{eq:G2_infinity} respectively as described above. Since each of these two pieces are obtained from adding terms corresponding to partitions of the two sets of space-time coordinates $\{x_1, ..., x_k\}$ and $\{y_1, ..., y_k\}$ (representing the coordinates of in-out $J^+$ and out-in $J^-$ sources respectively), the total number of such terms is equal to the product of two Bell numbers. For example, the averaged in-in two point correlator is given by
\bea
\overline{\langle\varphi(y_2) \varphi(y_1) \rangle^*\langle\varphi(x_2) \varphi(x_1)\rangle} = &\mathcal{W}^{\text{ii}}_{n^-_1n^+_1J^-_{1}J^-_{1}J^+_{1}J^+_{1}} +& \mathcal{W}^{\text{ii}}_{n^-_1n^+_2n^+_1J^-_{1}J^-_{1}J^+_{2}J^+_{1}} + \mathcal{W}^{\text{ii}}_{n^-_2n^-_1n^+_1J^-_{2}J^-_{1}J^+_{1}J^+_{1}} \notag\\
& + \mathcal{W}^{\text{ii}}_{n^-_2n^-_1n^+_2n^+_1J^-_{2}J^-_{1}J^+_{2}J^+_{1}}
\label{eq:Avg_ii_twopoint}
\eea
Since there are two terms for both in-out and out-in individually, there are four terms in total. 

It is worth mentioning that the generating functional for averaged in-out correlators $\mathcal{W}^{\text{io}}$ is contained within the generating functional for in-in correlators $\mathcal{W}^{\text{ii}}$ and thus the latter is a more general object. To get the former, one should not differentiate with respect to the minus sources and follow the regular procedure as described above for the averaged in-out correlators. Furthermore, the Euclidean version of this generating functional i.e. one with Euclidean fields, is straightforwardly obtained by doing a Wick rotation ($t \rightarrow -i\tau$ for '+' fields and $t \rightarrow i\tau$ for '-' fields). We now move on to real scalar field theory with a random mass and use the improved, more general generating functional \eqref{eq:G2_infinity} to calculate averaged in-out and in-in correlators. To illustrate the idea and be more simplistic, we do this for Euclidean scalar field. The results for Lorentzian/Quantum field theory can be easily obtained by undoing the Wick rotation in the end to obtain Quantum correlators from Euclidean correlators ($\tau \rightarrow i\,t$ and $\tau \rightarrow -i\,t$ for '+' and '-' fields respectively).

\section{Random Masses}\label{Sec:RandomMasses}

To illustrate the effect of random parameters in the action, we consider a massive scalar Euclidean field theory (Wick rotated Lorentzian theory) with source
\be
S_{E}(m)[\varphi, J] = \int d^4x_{E}\,\left(\frac{1}{2}\left(\partial\varphi\right)^2 + \frac{1}{2}m^2\varphi^2 - J\varphi\right)
\label{eq:Free_Action}
\ee 
where 
\be
(\partial\phi)^2 \equiv \left(\dfrac{\partial\phi}{\partial \tau}\right)^2 + \left(\dfrac{\partial\phi}{\partial x}\right)^2 + \left(\dfrac{\partial\phi}{\partial y}\right)^2 + \left(\dfrac{\partial\phi}{\partial z}\right)^2
\ee
with the following normalized distribution for mass
\be
P(m) = \dfrac{1}{\sigma\sqrt{2\pi}}\exp\left(-\dfrac{(m - m_o)^2}{2{\sigma}^2}\right) \label{eq:Mass_dist}.
\ee
The subscript E stands for Euclidean which we drop from now on assuming it is understood. Since the in-in generating functional \eqref{eq:G2_infinity} is a more general object, in the sense that it can generate both in-in and in-out correlators, we work with it (Euclidean version offcourse). Furthermore, to replicate all the partition functions in  \eqref{eq:G2_infinity} we assume that all $n$'s are positive integers. Accordingly, for a scalar field with action \eqref{eq:Free_Action} and distribution \eqref{eq:Mass_dist} of masses, the generating functional is given by
\be
\mathcal{W}^{\text{ii}} = \left [\int^{\infty}_{-\infty} \, dm \,P(m)\,\prod_{\alpha=1}^{\infty} \left (\prod_{\beta=1}^{n^+_\alpha} {Z_f}(m)[ J^+_{\alpha\beta}]\right) \left (\prod_{\gamma=1}^{n^-_\alpha} {Z_f}(m)[J^-_{\alpha\gamma}]\right)\right]_{J^+_{\alpha 1}=...=J^+_\alpha; \;J^-_{\alpha 1}=...=J^-_\alpha}
\label{eq:G2_mass}
\ee
where the second index for the sources (and fields) looks redundant, but is needed, because each of the partition functions (within a replicated set) is an integrated quantity and must be handled separately. Once we have calculated the partition functions, we drop the second subscript. To keep our notation simple, we don't bother writing $J^+_{\alpha 1}=...=J^+_\alpha; \;J^-_{\alpha 1}=...=J^-_\alpha$ anymore, with the understanding that it is implied. 

Since each of the partition functions are divergent quantities if directly calculated on a continuum space, we discretize Euclidean space-time and perform lattice field theory; the properly normalized correlators are obtained in the continuum limit. Coming back to \eqref{eq:G2_mass} with Eucledian space-time now discretized (i.e. coordinates $x$ being replaced with labels $k$ of lattice points, e.g. $\varphi^{\pm}_{\alpha\beta}(x)\rightarrow \varphi^{\pm}_{\alpha\beta,k}$), we separate the mass dependence from the partition functions and bring the integration over the mass inside the scalar fields' path integrals:
\bea
\mathcal{W}^{\text{ii}} = \int\prod_{\alpha=1}^{\infty} \left (\prod_{\beta=1}^{n^+_\alpha}D\varphi^+_{\alpha\beta} \prod_{\gamma=1}^{n^-_\alpha} D\varphi^-_{\alpha\gamma}\right)\left[\int^{\infty}_{-\infty} dmP(m)e^{-\frac{1}{2}m^2A}\right]\prod_{\alpha=1}^{\infty} \left (\prod_{\beta=1}^{n^+_\alpha}e^{-S(0)[ \varphi^+_{\alpha\beta}, J^+_{\alpha\beta}]}\prod_{\gamma=1}^{n^-_{\alpha}} e^{-S(0)[\varphi^-_{\alpha\gamma}, J^-_{\alpha\gamma}]}\right) \notag \\
\eea
 where
\be
A = \sum_{\alpha=1}^{\infty} \left(\sum^{n^+_{\alpha}}_{\beta=1} \sum_{k} \left(\varphi^+_{\alpha\beta,k}\right)^2 + \sum^{n^-_{\alpha}}_{\gamma=1} \sum_{j}\left(\varphi^-_{\alpha\gamma,j}\right)^2\right)
\label{eq:A}
\ee
After integration over the mass we get
\be
\mathcal{W}^{\text{ii}} =  \int\prod_{\alpha=1}^{\infty} \left (\prod_{\beta=1}^{n^+_\alpha}D\varphi^+_{\alpha\beta} \prod_{\gamma=1}^{n^-_\alpha} D\varphi^-_{\alpha\gamma}\right)\left[\dfrac{e^{-\frac{{m_o}^2A}{2}\left(1 + \sigma^2A\right)^{-1}}}{\sqrt{1 + \sigma^2A}}\right]\prod_{\alpha=1}^{\infty} \left (\prod_{\beta=1}^{n^+_\alpha}e^{-S(0)[\varphi^+_{\alpha\beta}, J^+_{\alpha\beta}]}\prod_{\gamma=1}^{n^-_{\alpha}} e^{-S(0)[\varphi^-_{\alpha\gamma}, J^-_{\alpha\gamma}]}\right).
\ee
Now to see what the total Euclidean action looks like, we move the term in square brackets into the exponent to give:
\bea
S_{\text{total}} = &\int d^4x\left[\sum_{\alpha=1}^{\infty}\left(\sum^{n^+_{\alpha}}_{\beta=1}\frac{1}{2}(\partial\varphi^+_{\alpha\beta})^2 - J_{\alpha\beta}^+\varphi_{\alpha\beta}^+ + \sum^{n^-_{\alpha}}_{\gamma=1}\frac{1}{2}(\partial\varphi^+_{\alpha\gamma})^2  - J_{\alpha\gamma}^-\varphi_{\alpha\gamma}^-\right)\right] \notag \\
&+ \dfrac{{m_o}^2A}{2}\left(1 + \sigma^2A\right)^{-1} + \dfrac{1}{2}\ln\left(1 + \sigma^2A\right)
\eea
with the understanding that we go back to continuum (Euclidean) space-time for the time being. 
Assuming $\sigma^2 \ll {m_o}^2$ and $m_o^2 A < 1$ (which will turn out to be equivalent to considering correlators at distances smaller than $1/m_o$) this can be expanded in powers of $\sigma^2/m_o^2$:
\be
\dfrac{{m_o}^2A}{2}\left(1 + \sigma^2A\right)^{-1} + \dfrac{1}{2}\ln\left(1 + \sigma^2A\right) = \dfrac{{m_o}^2\,A}{2} + \dfrac{\sigma^2}{2 m_o^2}\left(m_o^2 A - \dfrac{{m_o}^4}{2} A ^2\right)+ O\left (\dfrac{\sigma^4}{m_o^4} \right)
\ee
Note that the first term assigns a mass (equal to $m_o$) to all the fields and thus at this stage, we have a theory of infinite (replicated) massive fields, where non local interactions (coupled through powers of $\sigma^2/m_o^2$) will be treated perturbatively. Then, the free field operator for all the fields becomes $\Box + {m_o}^2$ and all the interaction terms can be recast by functional differentiation with respect to the sources as usual\footnote{It is worth mentioning that the \emph{effective} action for the scalar field would have $m_o$ as mass, which is clear from the construction of $\mathcal{W}$ as in the limit $\sigma \rightarrow 0$, the distribution in mass reduces to a delta distribution centered at $m_o$ which assigns mass equal to $m_o$ to all replicated fields with no interactions.}:
\be
\hat{A} = \sum_{\alpha=1}^{\infty} \left(\sum^{n^+_{\alpha}}_{\beta=1} \sum_{k} \left(\dfrac{\delta}{\delta J^+_{\alpha\beta,k}}\right)^2 + \sum^{n^-_{\alpha}}_{\gamma=1} \sum_{j} \left(\dfrac{\delta}{\delta J^{-}_{\alpha\gamma,j}}\right)^2 \right)
\ee
 Next, retaining only the leading order interaction term (corresponding to ${\sigma}^2/m_o^2$) in $\mathcal{W}^{\text{ii}}$ gives
\be
\mathcal{W}^{\text{ii}} \approx \left[1 - \dfrac{\sigma^2}{2 m_o^2}\left(m_o^2 \hat{A} - \dfrac{{m_o}^4}{2}\hat{A}^2\right)\right]\prod_{\alpha=1}^{\infty} \left (\prod_{\beta=1}^{n^+_\alpha} Z(m_o)[ J^+_{\alpha\beta}]\right) \left (\prod_{\gamma=1}^{n^-_\alpha} Z(m_o)[J^-_{\alpha\gamma}]\right)
\ee
After hitting with the operators $\hat{A}$, $\hat{A}^2$ and then collecting all similar terms, we can finally drop the second index introduced for the sources i.e. set $J^+_{\mu 1}=...=J^+_\mu; \;J^-_{\mu 1}=...=J^-_\mu$, to obtain
\bea
\mathcal{W}^{\text{ii}} & \approx \left[1 - \dfrac{\sigma^2}{2}\left(\sum_{\mu}n^+_{\mu}\dfrac{\sum_k\left(\frac{\delta}{\delta J^+_{\mu\,,k}}\right)^2Z[J^+_{\mu}]}{Z[J^+_{\mu}]}\right) + \dfrac{\sigma^2{m_o}^2}{4}\left(\sum_{\mu}n^+_{\mu}\Bigg\{\dfrac{\sum_k\sum_j\left(\frac{\delta}{\delta J^+_{\mu\,,k}}\right)^2\left(\frac{\delta}{\delta J^+_{\mu\,,j}}\right)^2Z[J^+_{\mu}]}{Z[J^+_{\mu}]} \right.\right.\notag\\
+& (n^+_{\mu}-1) \left(\dfrac{\sum_k\left(\frac{\delta}{\delta J^+_{\mu\,,k}}\right)^2Z[J^+_{\mu}]}{Z[J^+_{\mu}]}\right)^2\Bigg\} + \sum_{\mu,\nu}\dfrac{n^+_{\mu}n^+_{\nu}\left[\sum_k\left(\frac{\delta}{\delta J^+_{\mu\,,k}}\right)^2Z[J^+_{\mu}]\right]\left[\sum_j\left(\frac{\delta}{\delta J^+_{\nu\,,j}}\right)^2Z[J^+_{\nu}]\right]}{Z[J^+_{\mu}]\,Z[J^+_{\nu}]}\notag\\
+ &\{+ \leftrightarrow -\}+ 2\left.\left.\sum_{\mu,\nu}\dfrac{n^+_{\mu}n^-_{\nu}\left[\sum_k\left(\frac{\delta}{\delta J^+_{\mu\,k}}\right)^2Z[J^+_{\mu}]\right]\left[\sum_j\left(\frac{\delta}{\delta J^-_{\nu\,j}}\right)^2Z[J^-_{\nu}]\right]}{Z[J^+_{\mu}]Z[J^-_{\nu}]}\vphantom{1 +\,\sigma^2\sum_{\mu}\dfrac{n^+_{\mu}\sum_k\left(\frac{\delta}{\delta J^+_{\mu\,,k}}\right)^2Z[J_{\mu}]}{Z[J_{\mu}]}}\right)\right]\prod_{\mu=1}^{\infty} Z^{n^+_{\mu}}[ J^+_{\mu}]Z^{n^-_{\mu}}[J^-_{\mu}],\notag\\
\label{eq:W2_approx}
\eea
where we have suppressed the $m_o$ dependence of partition functions and $\{+ \leftrightarrow -\}$ represents the same terms as previous ones, but with time reversed (conjugated) objects \footnote{Note that in the Euclidean version, there is no difference between time and time reversed objects i.e. '+' and '-' partition functionals. But one should be careful as they are Wick (and reverse Wick) rotated differently to give different Lorentzian terms and hence we keep the subscripts differentiating the two types.}. Now it is apparent from the above equation \eqref{eq:W2_approx} that corrections (of order $O(\sigma^2/m_o^2))$ to averaged free in-out correlators originate from $\mathcal{W}^{\text{ii}}_{n_1}$ and $\mathcal{W}^{\text{ii}}_{n_1n_2}$ only. On the other hand, all of the possible partitions of the set of coordinates (as discussed in Sec. \ref{Sec:MultipleReplicas}) contribute non-trivially to averaged in-in correlators. Next we calculate the averaged in-out two point correlator and therefore only care about $\mathcal{W}^{\text{ii}}_{n_1}$ and $\mathcal{W}^{\text{ii}}_{n_1n_2}$:
\bea
\mathcal{W}^{\text{ii}}_{n_1} \approx \ln Z[J_1] -\dfrac{\sigma^2}{2}\sum_{k}\left(\Delta_{F,kk} + \left[\sum_l \Delta_{F,kl}J_{1,l}\right]^2\right) + \dfrac{\sigma^2{m_o}^2}{2}\sum_{k,j}\left(2\left[\sum_l \Delta_{F,kl}J_{1,l}\right]\Delta_{F,kj}\left[\sum^N_l \Delta_{F,jl}J_{1,l}\right]\right. \notag\\
+ \left.\left(\Delta_{F,kj}\right)^2\right);\notag
\eea
\be
\begin{split}
&\mathcal{W}^{\text{ii}}_{n_1\,n_2} \approx \ln Z[J_1]\ln Z[J_2] + \dfrac{\sigma^2{m_o}^2}{2}\left[\ln Z[J_1]\sum_{k,j}\Bigg\{2\left[\sum_l \Delta_{F,kl}J_{1,l}\right]\Delta_{F,kj}\left[\sum^N_l \Delta_{F,jl}J_{1,l}\right] + \Delta_{F,kj}\Delta_{F,kj}\Bigg\}\right.\\
& + \{1 \leftrightarrow 2\} + \left.\left(\sum_k\left(\Delta_{F,kk} + \left(\sum^N_j \Delta_{F,kj}J_{1,j}\right)^2\right)\right)\left(\sum_k\left(\Delta_{F,kk} + \left(\sum_j \Delta_{F,kj}J_{2,j}\right)^2\right)\right)\right]
\end{split}
\ee
As before, the subscripts for the free propagator $\Delta_{F}$ indicate lattice points ( i.e. $\Delta_{F,kj} \equiv \Delta_F(k - j)$). Note that since $A$ is quadratic, all odd point differentiations, i.e. differentiating an odd number of times with respect to sources, yield zero. Now differentiating with respect to sources, the averaged in-out two point correlator turns out to be
\bea
\overline{\langle\varphi(x_2)\varphi(x_1)\rangle}_{E} \approx \mathcal{W}^{ii}_{n_1 J_1 J_1} + \mathcal{W}^{ii}_{n_1 n_2 J_1 J_2} = \left(\begin{tikzpicture}[baseline=(current bounding box.center)]
\begin{scope}[very thick,decoration={
    markings,
    mark=at position 0.5 with {\arrow{>}}}
    ] 
\draw[postaction = {decorate}][gray, thick] (-0.5,0) -- (0.5,0);
\filldraw[black] (-0.5,0) circle (0pt) node[anchor=east] {$x_1$};
\filldraw[black] (0.5,0) circle (0pt) node[anchor=west] {$x_2$};
\end{scope}
\end{tikzpicture}\right)_{E}
- \sigma^2\left(\begin{tikzpicture}[baseline=(current bounding box.center)]
\begin{scope}[very thick,decoration={
    markings,
    mark=at position 0.5 with {\arrow{>}}}
    ] 
\draw[postaction = {decorate}][gray, thick] (-0.5,0) -- (0,0);
\filldraw[black] (-0.5,0) circle (0pt) node[anchor=east] {$x_1$};
\filldraw[black] (0,0) circle (1pt) node[anchor=south] {$z$};
\draw[postaction = {decorate}][gray, thick] (0,0) -- (0.5,0);
\filldraw[black] (0.5,0) circle (0pt) node[anchor=west] {$x_2$};
\end{scope}
\end{tikzpicture}\right)_{E} \notag\\
+ 4\sigma^2{m_o}^2\left(\begin{tikzpicture}[baseline=(current bounding box.center)]
\begin{scope}[very thick,decoration={
    markings,
    mark=at position 0.5 with {\arrow{>}}}
    ] 
\draw[postaction = {decorate}][gray, thick] (-0.5,0) -- (0.5,0);
\filldraw[black] (-0.5,0) circle (0pt) node[anchor=east] {$x_1$};
\filldraw[black] (0.5,0) circle (0pt) node[anchor=west] {$x_2$};
\filldraw[black] (-0.2,0) circle (1pt) node[anchor=south] {$z$};
\filldraw[black] (0.2,0) circle (1pt) node[anchor=south] {$z'$};
\end{scope}
\end{tikzpicture}\right)_{E}
\label{eq:avg_2p}
\eea
after taking the continuum limit of space-time. Note that only $\mathcal{W}^{\text{ii}}_{n_1}$ contributes to this averaged correlator. The subscript $E$ again stands for Euclidean and the diagrams are to be interpreted as usual with the identification of solid lines with massive ($m_o$) Euclidean propagators $\Delta_{F}$, dots with internal Euclidean space-time points that are to be integrated out, and arrows specifying with-time flow i.e. with usual propagators. The correction diagrams are standard integrals in Euclidean QFT \cite{6_Cardy,19_Ryder:1985wq} and are given as
\bea
\left(\begin{tikzpicture}[baseline=(current bounding box.center)]
\begin{scope}[very thick,decoration={
    markings,
    mark=at position 0.5 with {\arrow{>}}}
    ] 
\draw[postaction = {decorate}][gray, thick] (-0.5,0) -- (0,0);
\filldraw[black] (-0.5,0) circle (0pt) node[anchor=east] {$x_1$};
\filldraw[black] (0,0) circle (1pt) node[anchor=south] {$z$};
\draw[postaction = {decorate}][gray, thick] (0,0) -- (0.5,0);
\filldraw[black] (0.5,0) circle (0pt) node[anchor=west] {$x_2$};
\end{scope}
\end{tikzpicture}\right)_{E} = \dfrac{1}{(2\pi)^4}\int d^4k \dfrac{e^{i\,k^{\mu}x_{\mu}}}{(k^2 + m_o^2)^2} = \dfrac{1}{8\pi^2}K_{0}(m_o\,r)\notag\\
\left(\begin{tikzpicture}[baseline=(current bounding box.center)]
\begin{scope}[very thick,decoration={
    markings,
    mark=at position 0.5 with {\arrow{>}}}
    ] 
\draw[postaction = {decorate}][gray, thick] (-0.5,0) -- (0.5,0);
\filldraw[black] (-0.5,0) circle (0pt) node[anchor=east] {$x_1$};
\filldraw[black] (0.5,0) circle (0pt) node[anchor=west] {$x_2$};
\filldraw[black] (-0.2,0) circle (1pt) node[anchor=south] {$z$};
\filldraw[black] (0.2,0) circle (1pt) node[anchor=south] {$z'$};
\end{scope}
\end{tikzpicture}\right)_{E} = \dfrac{1}{(2\pi)^4}\int d^4k \dfrac{e^{i\,k^{\mu}x_{\mu}}}{(k^2 + m_o^2)^3} = \dfrac{r}{32\pi^2m_o}K_{1}(m_o\,r)
\eea
where $r$ is the Euclidean distance given as $\sqrt{(\tau_2 - \tau_1)^2 + (x_2 - x_1)^2 + (y_2 - y_1)^2 + (z_2 - z_1)^2}$ and $K_{n}(x)$ is the modified Bessel function of order $n$. Therefore in terms of Bessel functions, the averaged two point in-out correlator is approximately equal to 
\be
\overline{\langle\varphi(x_2)\varphi(x_1)\rangle}_{E} \approx \dfrac{1}{4\pi^2r^2}\left[m_orK_{1}(m_or) - \dfrac{\sigma^2}{2m_o^2}\left(m_o^2r^2K_{0}(m_or) - m_o^3r^3K_{1}(m_or)\right)\right].
\label{eq:avg_2p2}
\ee
 In the appendix \ref{Sec:Appendix}, we calculate the averaged in-out two-point Euclidean correlator by directly averaging the free massive Euclidean propagator over the distribution \eqref{eq:Mass_dist} and show that the two calculations  (leading to Eqs. \eqref{eq:avg_2p2} and \eqref{eq:avg_2p2_appendix} ) agree with each other to (at least) the first order in $\sigma^2/m_o^2$. Although this closed calculation can be done for random masses since there exists a closed form expression for in-out correlators for a free massive theory, it is not possible to do so in general. For instance if one is interested in having a distribution in the coupling constant, say $P(\lambda) \propto \lambda \exp\left( - \lambda^2/g^2 \right)$ for $\lambda \in [0;\infty)$, such that the standard perturbative analysis cannot be done for $\lambda > 1$, the direct averaging of the correlators is not possible. On the other hand for $g<1$ a perturbative expansion of our generating functionals can still be used for calculating (in general non-local) corrections to the averaged correlators in powers of $g$. However, to demonstrate that the above prescription works, we worked with only random masses for which a direct averaging of the correlator can be done to confirm its validity (See Appendix \ref{Sec:Appendix} for details).

At last, we find that the averaged in-in one-point correlator is zero as opposed to the case in Sec. \ref{Sec:GaussianField}.  To see this, we only need $\mathcal{W}^{\text{ii}}_{n^+_1n^-_1}$  which (upto order $\sigma^2/m_o^2$) is equal to
\bea
\mathcal{W}^{\text{ii}}_{n^+_1\,n^-_1} \approx \ln (Z[J^+_1]) \ln (Z[J^-_1]) + \dfrac{\sigma^2m_o^2}{2}\left[\ln (Z[J^-_1])\sum_{k,j}\Bigg\{2\left[\sum_l \Delta_{F,kl}J^+_{1,l}\right]\Delta_{F,kj}\left[\sum_l\Delta_{F,jl}J^+_{1,l}\right] \right.\;\;\;\;\;\;\;\;\;\;\\ \notag 
+ \Delta_{F,kj} \Delta_{F,kj}\Bigg\}+ \{+ \leftrightarrow -\} + \left.\sum_k\left(\Delta_{F,kk} + \left(\sum_j \Delta_{F,kj}J^+_{1,j}\right)^2\right)\times\sum_k\left(-\Delta_{F,kk} + \left(\sum_j \Delta_{F,kj}J^-_{1,j}\right)^2\right)\right]
\eea
and thus 
\be
\overline{\langle\varphi(y)\rangle^*\langle\varphi(x)\rangle}_{E} = \mathcal{W}^{\text{ii}}_{n^+_1 n^-_1 J^+_1 J^-_1} = 0
\ee
Although we do not show any calculations, for the averaged in-in two point correlator \eqref{eq:Avg_ii_twopoint} we need all four ($2 \times 2$) partitions i.e. $\mathcal{W}^{\text{ii}}_{n^+_1\,n^-_1}$, $\mathcal{W}^{\text{ii}}_{n^+_1\,n^-_1\,n^-_2}$, $\mathcal{W}^{\text{ii}}_{n^+_1\,n^+_2\,n^-_1}$ and $\mathcal{W}^{\text{ii}}_{n^+_1\,n^+_2\,n^-_1\,n^-_2}$; functional differentiation of which (with respect to sources at different points as discussed) would give the desired object.

\section{Discussion}\label{Sec:Discussion}

Quantum field theories with random potentials are of importance for condensed matter systems with the so-called quenched disorder  \cite{6_Cardy, 7_BDFN, 4_Anderson:1958vr, 5_Kamenev} and for cosmological systems in the context of string theory \cite{8_Tegmark:2004qd, 9_Denef:2004cf, 10_Marsh:2011aa}. In this paper we made the first step towards describing such systems using the method of generating functionals. More precisely, we constructed two different generated functionals for correlators in the case of a single quantum scalar field with random potentials using the replica trick. The first generating functional requires a single (double) set of replicated fields only, but involves taking an unphysical extrapolation to negative one number of fields (See Sec. \ref{Sec:SingleReplica}). The second generating functional has an infinite number of fields, but involves only an interpolation between zero and one number of fields in each replicated set (see Sec. \ref{Sec:MultipleReplicas}). The examples that we studied in detail were a random distribution of masses (see Sec. \ref{Sec:RandomMasses}) and two examples of random interactions, one described by an (Euclidean) action of the harmonic oscillator (see Sec. \ref{Sec:HarmonicOscillator}), and the other described by a Gaussian random potential (see Sec. \ref{Sec:GaussianField}). We have described how the perturbation theory for calculating in-out and in-in correlators works for these examples; the generalization to more fields and to different kinds of fields should be straightforward. We have also identified a technical problem for the distribution of random potential described by Euclidean field theories which limits our ability to perform perturbative calculations analytically, but can be dealt with numerically (see Sec. \ref{Sec:HarmonicOscillator}).  

Before we conclude let us briefly note that the probability of a particular type of scattering event is the square of the corresponding scattering matrix element which is an observable quantity. For example, considering a one to one scattering process for a real scalar field, the corresponding scattering amplitude $\bra{k}\ket{p}$ is related to the two point correlator $\langle\varphi(x_2)\varphi(x_1)\rangle$ through the LSZ formula \cite{18_Zee:2003mt, 19_Ryder:1985wq}, but the probability is given by $|\bra{k}\ket{p}|^2 $. Then the quenched averaged amplitude of such a scattering process is $\overline{ \bra{k}\ket{p}}$, but the averaged probability is $\overline{|\bra{k}\ket{p}|^2}$. For systems with random potentials, one can argue that we must not take quenched average over the correlators/Green's functions as they correspond to amplitudes for different potentials and taking averages as in \eqref{eq:quenched_io} would correspond to letting quantum trajectories on different potentials interfere which is unphysical. Thus, one should rather calculate the average of probabilities of different processes as in \eqref{eq:quenched_ii}. Although this is reasonable for condensed matter systems with quenched disorder, in quantum cosmology one might still argue that it makes sense for the wave function in different locations on the random landscape to interfere. To keep the discussion as general as possible we described the formalisms for calculating quenched averages of both amplitudes  \eqref{eq:quenched_io} and probabilities  \eqref{eq:quenched_ii}.  Application of the developed methods to inflationary cosmology will be discussed in a separate publication. \cite{33_Mudit_Vitaly}

{\it Acknowledgments.}  The authors are grateful to Thorsten Battefeld, Yi-Zen Chu and Alex Kamenev for very useful discussions and comments on the manuscript. The work was supported in part by Templeton Foundation and Foundational Questions Institute (FQXi).

\begin{appendix}

\section{Two-point correlator for a random mass}\label{Sec:Appendix}

Here we calculate the averaged two point in-out Euclidean correlator for random masses by directly averaging over the two point Euclidean correlator (Wick rotated Feynman propagator) for a massive free real scalar field which is given as 
\be
\Delta_{F}(x-y,m)_{E} = \dfrac{1}{(2\pi)^4}\int d^4k \dfrac{e^{i\,k^{\mu}x_{\mu}}}{(k^2 + m^2)} = \dfrac{1}{4\pi^2r^2}\int^{\infty}_0dz\,e^{(-z-\frac{m^2r^2}{4z})} = \dfrac{1}{4\pi^2r^2}\left(mrK_{1}(mr)\right)
\ee
With distribution \eqref{eq:Mass_dist} and working with the integral form for the propagator, we have  
\bea
\dfrac{1}{\sigma\sqrt{2\pi}}\int^{\infty}_{-\infty}\Delta_F(x_1 - x_2,m)_{E}e^{-\frac{(m - m_o)^2}{2\sigma^2}}dm = \dfrac{1}{4\sqrt{2}\pi^2\sqrt{\pi}r^2\sigma}\int^{\infty}_{0}dz\,e^{-z}\int^{\infty}_{-\infty}dm\,e^{-\frac{(m - m_o)^2}{2\sigma^2} - \frac{m^2r^2}{4z}}
\label{eq:avg_2p_full}
\eea
Completing squares in the gaussian integral for mass and then integrating it out gives
\be
\dfrac{1}{4\pi^2r^2}\int^{\infty}_0 dz \sqrt{\dfrac{z}{z + b^2}}e^{-z-\frac{a^2}{b^2 + z}}
\ee
where $a^2 = {m_o}^2r^2/4$ and $b^2 = \sigma^2r^2/2$. Change of variables $b^2 + z = x$ leads to the following integral
\be
\dfrac{e^{b^2}}{4\pi^2r^2}\int^{\infty}_{b^2} dx \sqrt{1 - \dfrac{b^2}{x}}e^{-x-\frac{a^2}{x}}
\ee
which, when expanded in powers of $b^2$ and retaining only first order terms in square root, gives
\be
\dfrac{e^{b^2}}{4\pi^2r^2}\int^{\infty}_{b^2} dx\left(1 - \dfrac{b^2}{2x}\right)e^{-x-\frac{a^2}{x}}
\ee
This can be recast as
\be
\dfrac{e^{b^2}}{4\pi^2r^2}\left[\int^{\infty}_{0} dx\left(1 - \dfrac{b^2}{2x}\right)e^{-x-\frac{a^2}{x}} - \int^{b^2}_{0} dx\left(1 - \dfrac{b^2}{2x}\right)e^{-x-\frac{a^2}{x}}\right]
\ee
Now for $b^2/a^2 \ll 1$ (which is equivalent to $\sigma^2/m_o^2 \ll 1$) the order of the second integral is bounded  by an exponentially small number $b^2 e^{(-a^2/b^2)}$ while the first term remains finite,  which leaves us with the first integral. Moreover we assume that $b^2 < 1$ (which is equivalent to considering distances $r < 1/m_o$) giving us
\be
\dfrac{1}{4\pi^2r^2}\left[\left(1 + b^2\right)\int^{\infty}_{0} dx\,e^{-x-\frac{a^2}{x}} - \dfrac{b^2}{2}\int^{\infty}_{0} dx\dfrac{e^{-x-\frac{a^2}{x}}}{x}\right]
\ee
to first order in $b^2$. This can be easily re-written as
\be
\dfrac{1}{4\pi^2r^2}\left[\left(1 + b^2\right)m_orK_{1}(m_or) + \dfrac{b^2}{2}\dfrac{\partial}{\partial (m_o^2r^2/4)}\left(m_orK_{1}(m_or)\right)\right]
\ee
which reduces to 
\be
\dfrac{1}{4\pi^2r^2}\left[m_orK_{1}(m_or) - \dfrac{\sigma^2}{2m_o^2}\left(m_o^2r^2K_{0}(m_or) - m_o^3r^3K_{1}(m_or)\right)\right]
\label{eq:avg_2p2_appendix}
\ee
and is the same as eq. \eqref{eq:avg_2p2}.

\end{appendix}

\end{document}